\documentclass[10pt, letter, twocolumn]{IEEEtran}

\usepackage{graphicx}
\usepackage{epsfig}
\usepackage{algorithm}
\usepackage{algorithmic}
\usepackage{ifmtarg}
\usepackage{caption}
\usepackage{footnote}
\usepackage{cite}
\usepackage{amssymb}
\usepackage{amsthm}
\usepackage[fleqn]{amsmath}
\usepackage{amsmath}
\usepackage{epstopdf}
\usepackage{amsfonts}
\usepackage{enumitem}
\usepackage{subfigure}
\usepackage{multirow}
\usepackage{rotating}
\usepackage{hyperref}
\usepackage{color}
\usepackage{comment}

\usepackage[font=normalsize]{caption}
\begin{document}
\title{Facilitating Satellite-Airborne-Balloon-Terrestrial Integration for Dynamic and Infrastructure-less Networks}
\author{\IEEEauthorblockN{Ahmad Alsharoa, \textit{IEEE Senior Member}, and Mohamed-Slim Alouini, \textit{IEEE Fellow}}
}

\maketitle
\thispagestyle{empty}
\pagestyle{empty}

\begin{abstract}
\boldmath{
This magazine investigates the potential enhancement of the data throughput of ground users by integrating ground base stations (GBS) with air stations, such as balloon, airborne, and satellite. The objective is to establish dynamic bi-directional wireless services (i.e., uplink and downlink) for ground users in congested and remote areas. The proposed integration involves satellite, high-altitude platforms (HAPs), and tethered balloons (TBs) in the exosphere, stratosphere, and troposphere, respectively, for better altitude reuse coupled with emerging optical or other high-frequency directional transceivers. This will lead to a signiﬁcant enhancement in scarce spectrum aggregate efﬁciency. However, the air stations deployment and resource managements in this integrated system 
faces difﬁculties. This article tackles resource management challenges by (i) providing wireless services to ground users in remote areas and connecting them with metropolitan and rural areas and (ii) employing HAPs equipped with free-space-optical communication modules as back-hauling backbones. Finally, we illustrate some numerical results to show the benefit of our proposed integrated system.
}
\end{abstract}


\section{Introduction and Motivation}\label{Introduction}

Satellite and ground base stations (GBSs), also known as `terrestrial stations’, are currently the main communication systems that provide wireless services to ground users in remote and metropolitan areas. While traditional space communications, including satellite stations, can deliver broadband services to ground users in remote areas, their spectral efﬁciency is constrained
because of the high pathloss attenuation of the channel between ground users and satellite station. Depending on satellite stations only can also cause extra delay for real time services because of their location at different orbital heights.

In contrast, GBSs cannot support ground users in remote areas due to their limited coverage areas and power unavailability. For this reason, one of the proposed ideas is to integrate the GBSs with satellite stations to improve the total network's capacity and coverage. Satellite stations can use multiple spot beams associated with multiple label switching protocols and spectrum access control~\cite{sat1}. Therefore, satellite stations can communicate with users with the help of GBSs working as relays~\cite{sat1}. In this case, GBSs are used to amplify the communication links between multiple satellite stations and multiple ground users~\cite{sat4}. However, because the satellite stations depend on the terrestrial network to broadcast their signals, this presents another limitation, especially in remote areas and during periods of congestion or network failure.


\noindent\textbf{High-altitude platforms (HAPs) and tethered balloons (TBs) can bridge this gap and provide downlink and uplink services in remote and congested areas.} One of the most important elements in the sixth-generation (6G) network is that coverage must be large enough to provide acceptable data-communication services wherever users live, including urban and remote areas. However, 6G networks are not intended to provide equally good service to all areas but rather maintain resource balance~\cite{6G}. Because of the huge growth in mobile and wireless device usage and immense data trafﬁc, traditional GBSs are expected to face some difﬁculties in supporting the demands of users in urban areas. This problem can be exacerbated by failures in the ground infrastructure. However, developing a terrestrial infrastructure that provides telecommunication services to remote areas is hardly feasible. To overcome these challenges, integrating the GBSs with higher altitude stations can be a promising solution for reaching global connectivity. 

The integration of GBSs, balloons, airborne, and satellite station into a single wireless network can enhance the overall network throughput. 
Further, the integration of the four stations would provide reliable, seamless, reasonable throughput anywhere on Earth. This includes remote, ocean, and mountain areas where the use of optical fiber is limited and costly. Indeed, nowadays only half the world’s population has access to the Internet according to recent statistics, and this unconnected population is mostly in the poorest areas of the world where infrastructure is scarce. 

HAPs and TBs can work as aerial relay stations to enhance the wireless channel quality between ground users and satellite station. Thus, they can enhance overall network throughput and help global connectivity with or without the existence of a terrestrial network\cite{HAP2}.
The altitude ranges of aerial relay stations are chosen carefully not only to reduce the needed consumed energy, but also to maintain the position stability of HAPs and TBs\cite{Airborne_survey}. This solution can provide immediate wireless connectivity to (i) ground users in remote areas with challenged networks, (ii) on-demand users in congested urban areas with capacity shortages due to peak trafﬁc, such as Olympic games, marathons, or base-station failures, (iii) ﬁrst responders and victims in emergency or disaster-recovery situations where infrastructure networks are unavailable or disrupted, (iv) ground military in hostile environments, and (v) border-patrol services for patrolling in difﬁcult terrain. The main advantages of deploying HAPs over GBSs can be summarized as\cite{Airborne_survey, HAP2}:
\begin{itemize}
  \item High coverage range: The GBSs' broadband coverage range is usually much less that the HAPs' range due to high non-line-of-sight (non-LoS) pathloss.
  \item Dynamic and quick deployment: HAPs have the flexibility of flying to remote or challenging areas to provide on-demand Internet services.
  \item Low consumed energy: Because of HAPs can be equipped with solar power panels, where the energy can be harvested during the daytime, then HAPs can be self-powered with careful trajectory optimization ~\cite{fb_HAP}.
\end{itemize}
In turn, the main advantages of HAPs over satellite stations can be summarized as~\cite{Airborne_survey}:
\begin{itemize}
  \item Quick and low-cost deployment: HAPs can be deployed quickly to accommodate traffic/temporal demand, emergency, or disaster-relief applications. One HAP being sufﬁcient to restart the broadband communications services by ﬂying to desired areas in a short, timely manner. Furthermore, The deployment cost of satellite station is much more than HAPs is HAPs' deployment cost.
  \item Low propagation delays and strong signals:  HAPs can provide broadband Internet services to ground users with less delay than satellite station. This is due to the lower pathloss attenuation compared to satellite stations.
\end{itemize}
Several papers have investigated the HAPs' deployment~\cite{HAP3,fb_HAP}. The authors in~\cite{HAP3} studied the deployment of the HAPs taking into account the ground users' quality of service (QoS).
The work, proposed a game theory model based on self organized model optimizes the
conﬁguration aiming to maximize the data rate of ground users. The work consider the HAPs as self organized and rational players.
While the authors in in~\cite{fb_HAP} proposed some techniques for trajectory optimization where HAPs are equipped with solar panels. The work proposed two solutions based on heuristic greedy algorithm and real-time and low complexity solution. The objective is to minimize the consumed energy, that is constrained by the harvested energy amount, by optimizing the trajectory of the HAPs. Improving the overall system data rate is also limited by another key factor in this integration system. For instance, the authors in~\cite{HAP4} proposed to use orthogonal frequency division multiple access (OFDMA) for multicast technique
to optimize the HAPs' transmit powers, transmission time slots, and resource block channels aiming to maximize the total ground user data rate. In other words, they maximized the number of users that received the requested multicast streams within the HAP's service area for a given OFDMA slot frame. The achieved enhancement in multiple HAPs’ capacities is discussed in~\cite{HAP5}, where the authors showed that by exploiting the directionality of user antennas, HAPs can offer spectral efficiency. This work is also discussed 
how multiple HAPs can take the advantages of directionality by sharing the same frequency bandwidth. In a recent work of ours~\cite{Alsharoa_J16}, we proposed a downlink resource-allocation solution of integrated satellites, HAPs, and GBSs based on OFDMA to maximize user throughput considering back-hauling and access-link constraints. In another work~\cite{Alsharoa_J14}, we proposed using TBs connected with optical ﬁbers as back-hauling stations to support ground users.

\noindent\textbf{Equipping aerial relay stations with free-space-optical (FSO) transceivers is a tipping point.} Limited research has proposed equipping aerial relay stations (i.e., HAPs and TBs) with FSO transceivers ~\cite{FSO10,FSO17}. The overview of equipping HAPs with optical transmitters has been discussed in~\cite{FSO10}. The work shows that several Gbps data throughput can be achieved by using leaser FSO beams.  In~\cite{FSO17}, in contrast, closed form expressions for bit-error-rate and average capacity were derived using multiple hop FSO transmitters in the stratosphere region. However, all the previous works did not consider managing the resource allocations in a satellite-airborne-balloon-terrestrial network integration while maintaining the FSO communication links, nor did they consider the issues of back-hauling and access-link communications.

\section{Integration with hybrid FSO/RF}

\subsection{Integration of Aerial Relay Stations}
Deployment of aerial relay stations in remote, large geographical infrastructure-less areas and the regions suffer from grid power limitation can be a potential solution to provide wireless uplink and downlink services to ground users.
Working as relays or intermediate nodes, aerial stations can maintain the wireless connectivity by broadcasting the uplink and downlink signals.
aerial relay stations not only enhance the uplink and downlink signals but also create a new dimension to next-generation wireless networks and service provisioning. For example, in September 2017, Hurricane Irma (Category 4 Hurricane) hit Florida and damaged a significant percentage of cellular GBSs. For example, in some counties in Miami city, Irma caused more than 50\% of GBS failure even few days after the hurricane. Additionally, Irma knocked out power to more than 6.8 million people for several days/weeks~\cite{FCCirma}.
In such circumstances, aerial relay stations can reach such affected areas because of their quick and dynamic deployment~\cite{2017-Stumpe-drones}. 

\begin{figure}[h!]
\includegraphics[width=3.5in]{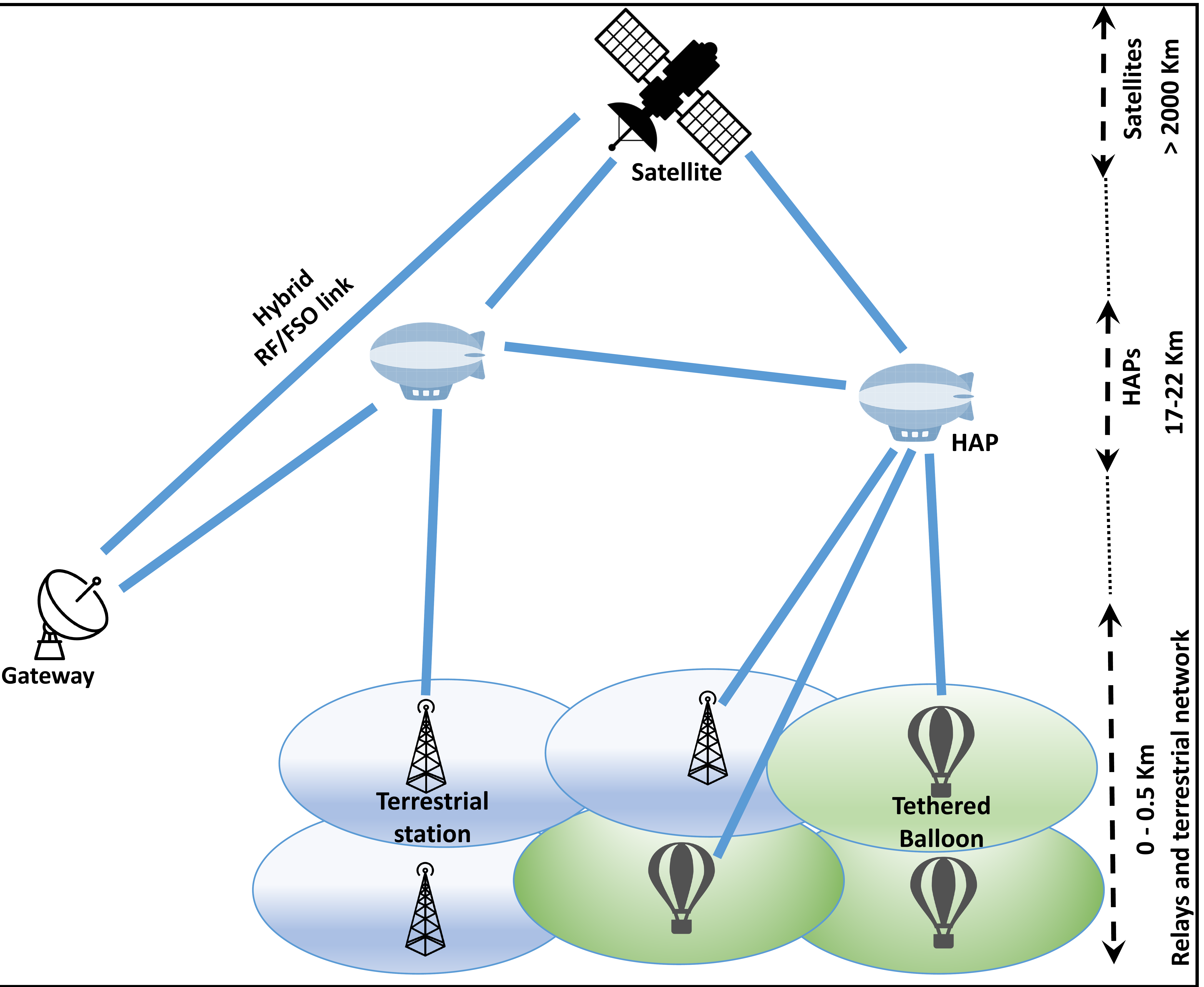}
\caption{System model.}\label{SystemModel_u3}
\end{figure}

The success of deploying aerial relay stations in remote or challenging areas depends mainly on two factors. The first factor is the integration with ground users via access link. While the second factor is the availability and other parameters related to back-hauling links.
The efficient placement and resource managements of aerial relay stations wonderfully can generates wireless connectivity in challenging areas. Note that, TBs can be powered by renewable energy (RE) sources, and therefore, can be placed in remote areas to help in uplink and downlink services. This is consider as a plausible proposal since in remote areas the TBs would consume less energy by serving fewer users. Additionally, some sleeping strategy (i.e., powering on-off the TBs) can be used to reduce the overall energy consumption.

\subsection{Hybrid FSO/RF}
Compared to the current wireless networks, the expected next generation wireless demand is ambitious and may require to support throughput around 1,000 times higher with round-trip latency around 10 times lower ~\cite{5Gpaper}. The radio frequency (RF) spectrum is anticipated to be more and more congested for emerging technologies and applications in future wireless networks. Therefore, the RF technology maybe insufficient to accommodate the expected increase in the demand of wireless devices. Focusing on improving the RF spectrum only in legacy bands may be not enough, thus it is critical to embrace a comprehensive technology with \emph{high spectral reuse} by supplementing RF technology with other wireless technologies in directional higher frequency bands~\cite{2013-Sevincer-LIGHTNETS}. FSO communications is considered as a promising complementary solution with 
RF to meet the exploding demand for wireless networks. FSO transceivers are amenable to dense integration and provide spatial reuse and security through highly directional beams. RF’s huge unlicensed spectrum presents a unique opportunity for to overcome the expected future spectrum scarce problem. Its potential integration with solid-state lighting technology also presents an attractive commercialization possibility~\cite{2013-Sevincer-LIGHTNETS}. The authors in~\cite{FSO_5G} proposed a vertical framework consisting of networked HAPs that supported back-hauling links and access links of small cell base stations in a multi-tier heterogeneous network. However, that work was limited to supporting small cells, did not consider the integration of all types of GBSs (e.g., macro cell base stations) and satellite stations with HAPs, and did not discuss providing global connectivity in remote areas. This paper focuses on infrastructure-less operation, where the HAPs and TBs are equipped with directional FSO transceivers to provide wireless connectivity for the back-hauling links.

\begin{figure}[t!]
\includegraphics[width=3.5in]{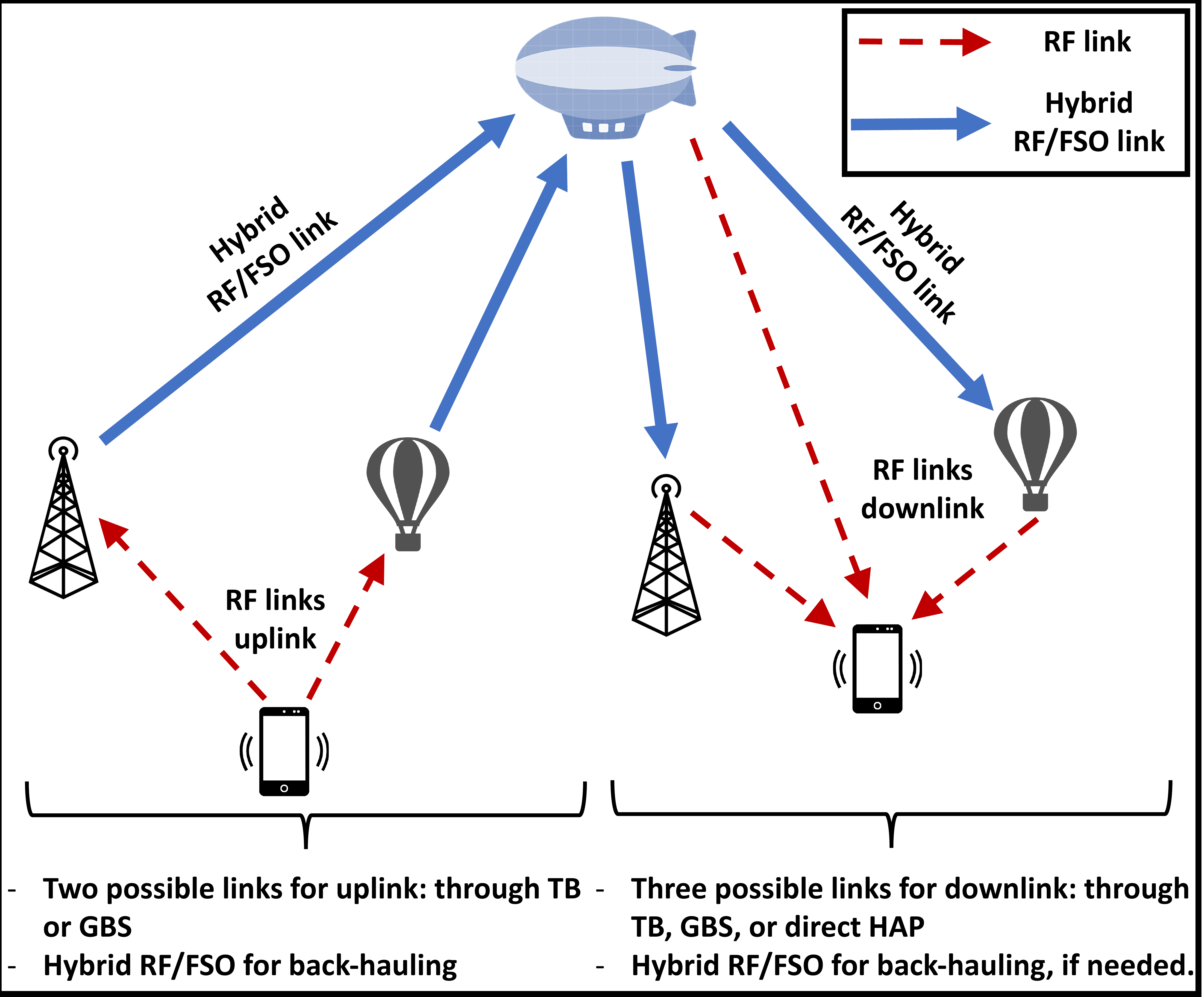}
\caption{All possible uplinks and downlinks scenarios}\label{4scenarios}
\end{figure}

HAP-TB integration offers another dimension to legacy wireless networking by enabling spatial reuse. TBs can effectively amplify the signals between the HAPs and ground users using FSO links without causing any major interference for the rest of the ground users. If this spatial technique is not used and relaying is used only on the RF band, then the aggregated throughput will be limited due to (i) the scarcity of the RF band and (ii) the possibility of interference between HAPs, TBs, and ground users.
Therefore, the potential of using spatial reuse is possible only if HAPs, TBs, and gateways are equipped with FSO directional antennas and their positioning and precision steering of the antennas are feasible. However, the FSO directional transceivers requires not only the FSO link establishment, but also line-of-sight maintenance between different transceivers.

\section{System Model}
We deﬁne the following system model as a hybrid FSO link:
\begin{itemize}[leftmargin=*]
\item
A set of stations, including a satellite, HAPs, TBs, and GBSs, where each station contains its station ID, 3D location, battery level at any given time (if applicable, e.g., HAPs and TBs), and back-hauling rate. All this information is shared with some central units. Since the TBs are RE stations, the consumed power of the TBs, such as operating and transmission, is included.
\item
A set of ground users that contain user IDs, 2D locations, and QoS requirements.
\item
A set of ﬁx gateways that contain the gateway IDs and 2D locations.
\end{itemize}

The major challenge is managing the ground users' and stations' resources stations is to maximize the ground users’ data rate considering the (i) bandwidth and power constraints, (ii) association constraints, and (iii) HAP- and TB-placement constraints. In other words, the ground users’ throughput depends on several factors such as the maximum ground users' allowable transmit power, TBs' and HAPs' available bandwidth, and station placement. Therefore, control links between the stations and corresponding users are required to allow stations to track users’ locations under its coverage area and manage resources. The rate utility of the system can be characterized by various utility metrics, the selection of which can be based on required fairness levels. Some examples are (i) sum rate utility (maximize the sum rate throughput of all users), (ii) minimum rate utility (maximize the minimum user throughput), and (iii) proportionally fair rate utility (maximize the geometric mean of the data rate).

\section{Resource Management}
The key  goal is to attain high throughput of ground users and energy efficiency. Several metrics can be implemented to achieve this goal, such as (1) minimizing the total consumed energy while satisfying a certain user's throughput, or (2) maximizing the energy efficiency utility.
In this context, several high-level research questions need to be addressed:
\begin{itemize}
    \item \textit{Resource Optimization}: How to optimize the transmit power allocation of the users and various types of stations and, given a certain available bandwidth, how to allocate this bandwidth for the control links (for management) or serving links (i.e., access and back-hauling links)
    \item \textit{Associations}: (a) Access link associations (the associations between users and GBSs, TBs, and HAPs) and (b) the back-hauling link associations (the associations of GBSs and TBs with HAPs and between HAPs and gateways), as shown in Fig.~\ref{4scenarios}.
    \item \textit{HAPs and TBs Placement}: How to ﬁnd the best locations for HAPs and TBs considering back-hauling link quality.
    \item \textit{FSO Alignment}: How to optimize the FSO alignment angles between different FSO transceivers.
\end{itemize}

The RF and FSO choices will depend on several factors, such as environmental or weather conditions and the feasibility of LoS. Note that the use of a hybrid FSO/RF link will be for back-hauling links, while the RF band only can be used for the access link due to the difﬁculties of tracking the movement of ground users and maintaining the LoS, as shown in Fig.~\ref{4scenarios}.

\subsection{Access Link Optimization}
In this work, we propose using multiple stations (i.e., GBSs, RE TBs, and HAPs) to provide wireless connectivity to multiple ground users. Because we involve different types of stations, the access link can be established based on the form of communications as follows: a) For uplink, two possible links can be established: ground users to GBS or ground users to TBs. In contrast, b) for downlink, three possible links can be established: HAP to ground users, GBS to ground users, or TB to ground users, as shown in the dashed lines in Fig.~\ref{4scenarios}. In this case, the mathematical formulation should include an access link binary variable to indicate that ground users are associated with certain stations for the speciﬁc form of communications (either downlink or uplink). For simplicity, we assume that each user can be associated with one station at most; however, each station can be associated with multiple users. For station peak power and user peak power, an optimization problem can be formulated that maximizes user utility given the following constraints: (1) back-hauling bandwidth and rate, (2) station and user peak powers, (3) access link associations, and (4) user QoS. Therefore, the following parameters can be optimized to achieve the best objective function: (i) the transmit power levels of the user and station transmission power, (ii) bandwidth allocation to each user, and (iii) access link associations.

Another two factors can play signiﬁcant roles in determining the access link associations: first, the back-hauling available rates, where a user can be associated with a distant station if it has a good back-hauling rate rather than being associated with a nearby station with a low back-hauling rate, and second, the energy stored in TBs because they are assumed to be RE battery-powered stations and the amount of stored energy by each TB at the end of a given time slot is considered an additional limitation. In this case, each TB should respect and ensure that the consumed energy is less than the stored energy in the previous time slot.

\subsection{Back-hauling Optimization}
In this section, we propose integrating GBSs and TBs with HAPs using hybrid FSO/RF links. A key challenge for networking under partial or no infrastructure is establishing reliable back-hauling links to the TBs or gateways involved in the provisioning of connectivity services for ground users. The back-hauling optimization problem is proposed to optimally ﬁnd back-hauling associations, HAP locations, transmit powers, and FSO alignment between transceivers to maximize user back-hauling throughput while respecting resource limitations. This involves utilizing high-frequency directional bands, such as optical bands, as much as possible to minimize the interference between transceivers.
It is assumed that each HAP is strictly associated to one back-hauling station (i.e., either a gateway station or satellite station). additionally, we assume a limited number of HAPs associated with the same back-hauling station. For the FSO link between different transceivers, we assume that the alignment angle can be optimized to achieve LoS alignment. In this case, we propose FSO link discovery and establishment. One way is to explore out-of-band techniques where support from an RF or GPS is available to exchange the angle/direction and also exchange information about how they are oriented.

In addition to the HAP back-hauling, back-hauling links between GBSs and TBs with HAPs can be established based on the form of communications as follows: a) for uplink, two possible links can be established: GBS to HAPs or TBs to HAPs; b) for downlink, two possible links can be established: HAP GBSs or HAPs to TBs, as shown in the solid lines in~\ref{4scenarios}. Therefore, another binary decision variable must be introduced for the back-hauling associations.

\section{Results and Discussion}
\begin{figure}[b!]
\includegraphics[width=3in]{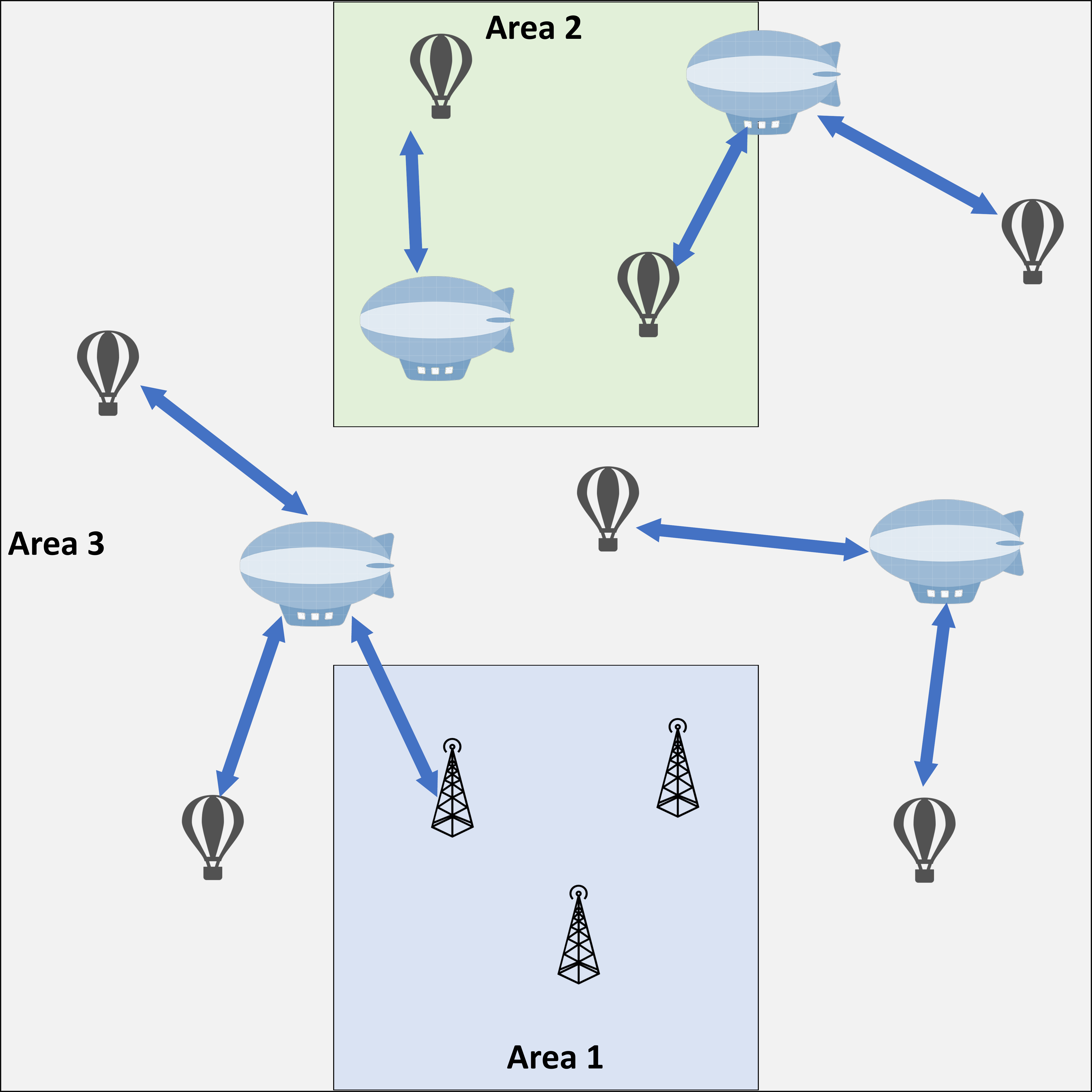}
\caption{Simulation Setup}\label{sim1}
\end{figure}

This section provides some numerical results to outline the beneﬁts of using our proposed integration to improve global connectivity. The numerical results setup is within an area of 180km $\times$ 180km. We distribute $U$ ground users in this area in three different sub-areas: i) sub-area A: 70km $\times$ 70km, ii) sub-area B: 70km $\times$ 70km, and iii) sub-area C: the remaining area. We also consider different users' density distributions in each sub-area, as shown in Fig.~\ref{sim1}. Sub-area A contains 30 GBSs with x and y ranges as (x:55–125 km) and (y: 0–70 km), respectively, and consists 40\% of total number of ground users. Subarea B has no GBSs with x and y ranges as (x:55–125 km) and  (y:110–180 km), respectively, and consists of 30\% of total number of ground users. Sub-area C is the remaining area. Further, the total number of TBs and HAPs used are 30 and 8, respectively, as given in Table.~\ref{tab1}. 

We study the enhancement of the achievable downlink and uplink throughput when HAPs and TBs assist the terrestrial network. We also consider different back-hauling bandwidth cases to represent both RF-only and hybrid FSO/RF scenarios to investigate the limitation of the RF-only scenario.

{
\begin{table}[h]
\centering
\caption{\label{tab1} Simulation parameters}
\addtolength{\tabcolsep}{-2pt}\begin{tabular}{|l||c|c|c|}
\hline
\textbf{} & \textbf{Sub-area A} & \textbf{
Sub-area B} & \textbf{Sub-area C} 
\\ \hline \hline
Sub-area (Km $\times$ Km) & 70 $\times$ 70 & 70 $\times$ 70 & Remaining  
\\ \hline
Dimensions (Km) & [x:55-125], & [x:55-125], & Remaining  
\\
 & [y:0-70] & [y:110-180] & {}
\\ \hline
Users distribution (\%)
& 40 & 30  & 30
\\ \hline
Number of TBSs
 & 30 & 0 & 0 
\\ \hline
Example
 & City area
 & After disaster area & Rural area
\\\hline
\end{tabular}
\end{table}
}

The results in this section are based on optimizing the station placement (i.e., HAPs and TBs) and the associations (access and back-hauling associations) as follows: (i) stations placements: we use a heuristic shrink-and-re-align algorithm to ﬁnd the optimal placement. The shrink-and-re-align algorithm has several benefits compare to other heuristic algorithms in the literature: (i) it is has a simple implementation, (ii) considered as low complexity algorithm, and (ii) very fast and quick convergence. The algorithm starts by generating initial next candidate positions as spheres around the current position before recursively shrinking the sphere radius by half to ﬁnd the best local position and compare it the current position. The algorithm repeats the above process until the sample space size decreases below a specific limit or no enhancement can be made.
ii) Resource allocations: the resource allocations are optimizing based on OFDMA, we formulate a linear and convex optimization problems for solving the associations and power allocations.

\begin{figure}[h!]
\includegraphics[width=3.4in]{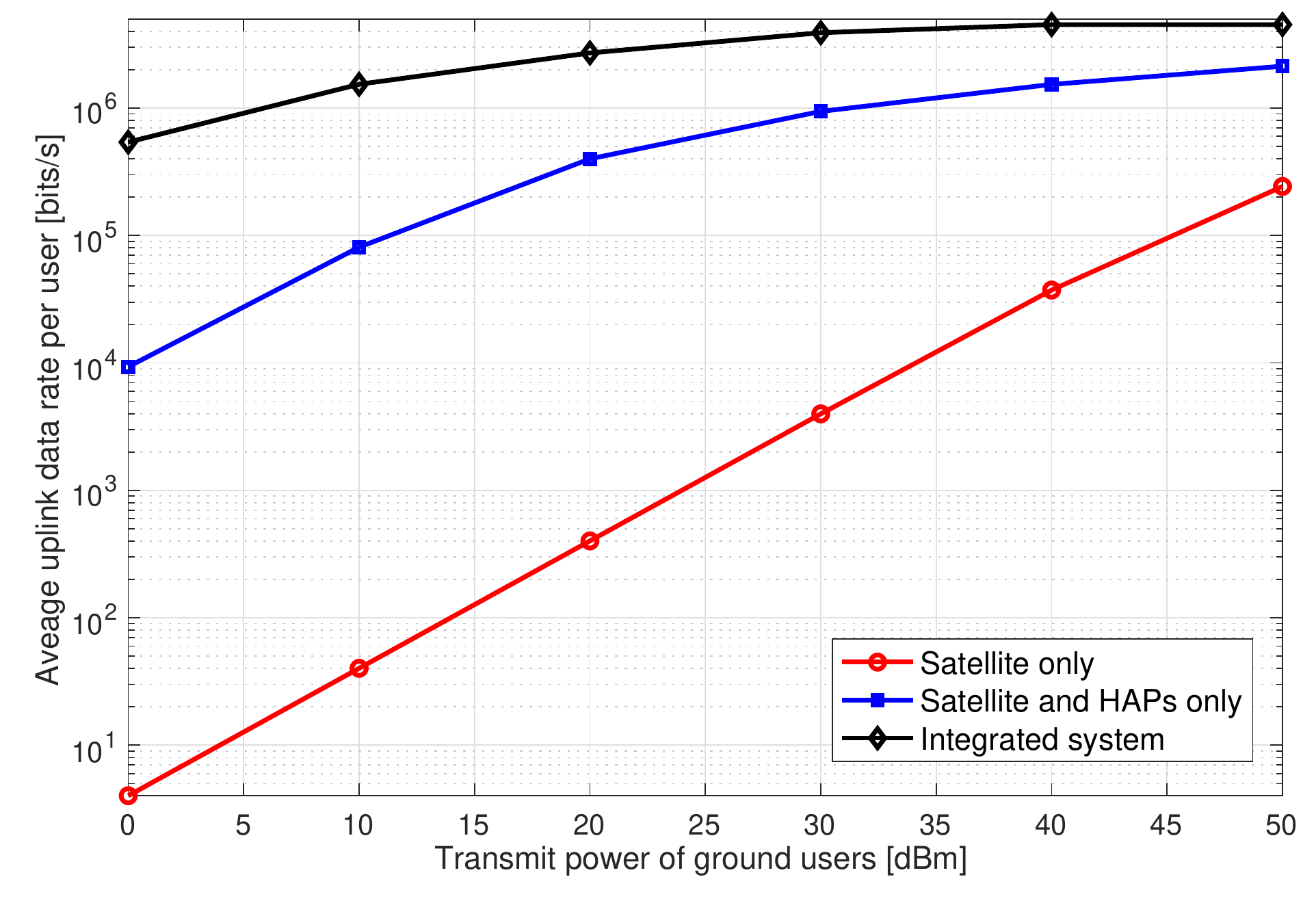}
\caption{Average uplink data rate versus ground user's transmit power.}\label{fig1}
\end{figure}

Fig.~\ref{fig1} plots the achievable average uplink data rate as a function of transmit power of ground user. The average uplink data rate is per ground user and equals to 
the sum of the total uplink throughput divided by the total number of ground users. This figure compares our proposed integration model with two baseline models: 1) satellite only: the satellite station is lonely used for the uplink transmission, 2) satellite and HAPs only: satellite and multiple HAPs are used only (i.e., TBs are excluded) for the uplink transmission. It is shown that our integrated system achieves a higher uplink throughput compares to the two baselines.
For example, when the ground user transmitted power is equal to 20dBm, the achievable uplink throughput can be improved from around 0.3 Kbits/sec for satellite only solution and 0.3 Mbits/sec for satellite and HAPs only to around 3 Mbits/sec by using the proposed integrated system.
This is due to the help of TBs that can help in broadcasting the uplink signal and mitigating the pathloss and other unfavorite effects. Further, from the same figure we can show that the average ground users throughput of proposed integrated system and satellite and HAPs only increases as the ground transmit power increases up to a specific value ($\approx 40$ dBm) 
After this value, the performance remains constant. This is because of the back-handing limitation in the relay link from TBs to HAPs.

\begin{figure}[h!]
\includegraphics[width=3.4in]{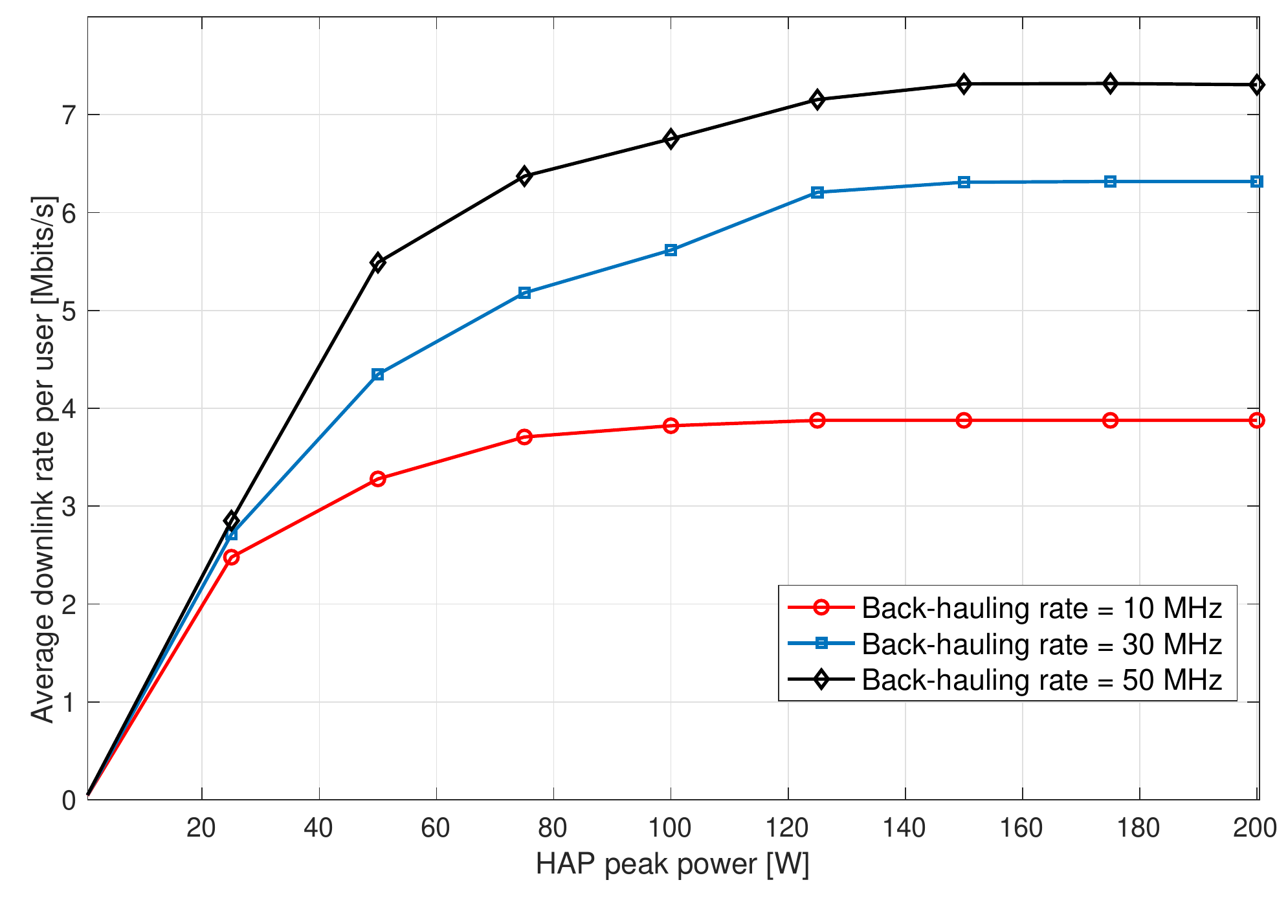}
\caption{Average downlink data rate versus HAPs' peak power.}\label{fig3P}
\end{figure}

To illustrate the FSO/RF back-hauling bottleneck effects, Fig.~\ref{fig3P} illustrates the average downlink throughput as a function of HAPs peak transmit power. 
It can be seen that as the downlink data rate increases as HAPs’ peak transmit power rises up to a certain value.
This is because, by starting from this value, the average downlink throughput can not be improved. This is due to the limitation of the back-hauling link from HAPs to gateways or HAPs to satellite links. Further, it can be seen that the average downlink throughput increases as back-hauling bandwidth raises. This is because increasing the back-hauling bandwidth also increases the back-hauling data rate capacity, thus increasing the back-hauling bottleneck. Therefore, hybrid FSO/RF communication links can be used to mitigate back-hauling bottleneck limitations and thus enhance the overall performance.


\section{Conclusion}
This paper proposes an efﬁcient scheme that integrates GBSs, TBs, HAPs, and satellite stations to provide global connectivity. Our objective is to improve downlink and uplink rates by optimizing back-hauling and access links. We also proposed equipping the stations with FSO transceivers to mitigate the back-hauling bottleneck limitation as shown in the simulation result section and therefore improve the data rate. In contract, this needs more effort in optimizing extra variables such as line-of-sigh angels. 

\bibliographystyle{IEEEtran}
\bibliography{CM_2019HAP_15ref}

\begin{thebibliography}{10}
\providecommand{\url}[1]{#1}
\csname url@samestyle\endcsname
\providecommand{\newblock}{\relax}
\providecommand{\bibinfo}[2]{#2}
\providecommand{\BIBentrySTDinterwordspacing}{\spaceskip=0pt\relax}
\providecommand{\BIBentryALTinterwordstretchfactor}{4}
\providecommand{\BIBentryALTinterwordspacing}{\spaceskip=\fontdimen2\font plus
\BIBentryALTinterwordstretchfactor\fontdimen3\font minus
  \fontdimen4\font\relax}
\providecommand{\BIBforeignlanguage}[2]{{%
\expandafter\ifx\csname l@#1\endcsname\relax
\typeout{** WARNING: IEEEtran.bst: No hyphenation pattern has been}%
\typeout{** loaded for the language `#1'. Using the pattern for}%
\typeout{** the default language instead.}%
\else
\language=\csname l@#1\endcsname
\fi
#2}}
\providecommand{\BIBdecl}{\relax}
\BIBdecl

\bibitem{sat1}
X.~{Yan}, H.~{Xiao}, C.~{Wang}, and K.~{An}, ``Outage performance of
  {NOMA}-based hybrid satellite-terrestrial relay networks,'' \emph{IEEE
  Wireless Communications Letters}, vol.~7, no.~4, pp. 538--541, August 2018.

\bibitem{sat4}
U.~{Park}, H.~W. {Kim}, D.~S. {Oh}, and B.~J. {Ku}, ``Flexible bandwidth
  allocation scheme based on traffic demands and channel conditions for
  multi-beam satellite systems,'' in \emph{Proc. of the IEEE Vehicular
  Technology Conference, Quebec City, QC, Canada}, September 2012, pp. 1--5.

\bibitem{6G}
S.~{Dang}, O.~{Amin}, B.~{Shihada}, and M.-S. {Alouini}, ``From a human-centric
  perspective: What might {6G} be?'' \emph{Nature Electronics}, vol.~3, no.~1,
  pp. 20--29, January 2020.

\bibitem{HAP2}
H.~{Lu}, Y.~{Gui}, X.~{Jiang}, F.~{Wu}, and C.~W. {Chen}, ``Compressed robust
  transmission for remote sensing services in space information networks,''
  \emph{IEEE Wireless Communications}, vol.~26, no.~2, pp. 46--54, Apr. 2019.

\bibitem{Airborne_survey}
X.~{Cao}, P.~{Yang}, M.~{Alzenad}, X.~{Xi}, D.~{Wu}, and H.~{Yanikomeroglu},
  ``Airborne communication networks: {A} survey,'' \emph{IEEE Journal on Sel.
  Areas in Communications}, vol.~36, no.~9, pp. 1907--1926, September 2018.

\bibitem{fb_HAP}
J.~{Marriott}, B.~{Tezel}, Z.~{Liu}, and N.~{Stier}, ``Trajectory optimization
  of solar-powered high-altitude long endurance aircraft,'' \emph{Facebook
  Research}, October 2018.

\bibitem{HAP3}
R.~{Zong}, X.~{Gao}, X.~{Wang}, and L.~{Zongting}, ``Deployment of high
  altitude platforms network: {A} game theoretic approach,'' in \emph{Proc. of
  the International Conference on Computing, Networking and Communications,
  Maui, HI, USA}, January 2012, pp. 304--308.

\bibitem{HAP4}
A.~{Ibrahim} and A.~S. {Alfa}, ``Using {L}agrangian relaxation for radio
  resource allocation in high altitude platforms,'' \emph{IEEE Tran. on
  Wireless Communications}, vol.~14, no.~10, pp. 5823--5835, October 2015.

\bibitem{HAP5}
D.~{Grace}, J.~{Thornton}, {Guanhua Chen}, G.~P. {White}, and T.~C. {Tozer},
  ``Improving the system capacity of broadband services using multiple
  high-altitude platforms,'' \emph{IEEE Tran. on Wireless Communications},
  vol.~4, no.~2, pp. 700--709, March 2005.

\bibitem{Alsharoa_J16}
A.~{Alsharoa} and M.~{Alouini}, ``Improvement of the global connectivity using
  integrated satellite-airborne-terrestrial networks with resource
  optimization,'' \emph{IEEE Transactions on Wireless Communications}, vol.
  {Early Access}, 2020.

\bibitem{Alsharoa_J14}
A.~{Alzidaneen}, A.~{Alsharoa}, and M.~{Alouini}, ``Resource and placement
  optimization for multiple {UAVs} using backhaul tethered balloons,''
  \emph{IEEE Wireless Communications Letters}, vol.~9, no.~4, pp. 543--547,
  April 2020.

\bibitem{FSO10}
F.~{Fidler}, M.~{Knapek}, J.~{Horwath}, and W.~R. {Leeb}, ``Optical
  communications for high-altitude platforms,'' \emph{IEEE Journal of Sel.
  Topics in Quantum Electronics}, vol.~16, no.~5, pp. 1058--1070, September
  2010.

\bibitem{FSO17}
M.~{Sharma}, D.~{Chadha}, and V.~{Chandra}, ``High-altitude platform for
  free-space optical communication: Performance evaluation and reliability
  analysis,'' \emph{IEEE Journal of Optical Communications and Networking},
  vol.~8, no.~8, pp. 600--609, August 2016.

\bibitem{FCCirma}
{Federal Communications Commission}, ``Communications status report for areas
  impacted by hurricane {I}rma {S}eptember 11, 2017,'' {FCC report}, Sept.
  2017,
  \url{https://www.fcc.gov/document/hurricane-irma-communications-status-report-sept-11}.

\bibitem{2017-Stumpe-drones}
J.~Stumpe, ``Drones to the rescue: Airline industry embraces drones as
  cost-saver,'' {Aerospace America}, November 2017,
  \url{https://aerospaceamerica.aiaa.org/features/drones-to-the-rescue}.

\bibitem{5Gpaper}
J.~G. {Andrews}, S.~{Buzzi}, W.~{Choi}, S.~V. {Hanly}, A.~{Lozano}, A.~C.~K.
  {Soong}, and J.~C. {Zhang}, ``What will {5G} be?'' \emph{IEEE Journal on
  Selected Areas in Communications}, vol.~32, no.~6, pp. 1065--1082, June 2014.

\bibitem{2013-Sevincer-LIGHTNETS}
A.~Sevincer, A.~Bhattarai, M.~Bilgi, M.~Yuksel, and N.~Pala, ``{LIGHTNETs:}
  smart lighting and mobile optical wireless networks -- a survey,'' \emph{IEEE
  Comm. Surveys \& Tutorials}, vol.~15, no.~4, pp. 1620--1641, April 2013.

\bibitem{FSO_5G}
M.~{Alzenad}, M.~Z. {Shakir}, H.~{Yanikomeroglu}, and M.-S. {Alouini},
  ``{FSO}-based vertical backhaul/fronthaul framework for {5G+} wireless
  networks,'' \emph{IEEE Communications Magazine}, vol.~56, no.~1, pp.
  218--224, Jan. 2018.

\end{thebibliography}

\end{document}